\documentclass[12pt,a4paper]{article}
\usepackage{amsfonts}
\usepackage{amssymb}
\usepackage{indentfirst}
\usepackage{graphics}
\usepackage{graphicx}
\usepackage{epsfig}

\newcommand\be{\begin{equation}}
\newcommand\ee{\end{equation}}
\renewcommand{\theequation}{\thesection.\arabic{equation}}

\begin{document}

\title{The Unique Determination of Neuronal Currents in the Brain
via Magnetoencephalography}
\author{A.S.\ Fokas$^\dag$, Y.\ Kurylev$^\ddag$ and
V.\ Marinakis$^\dag$ \\
\\
$^\dag${\em Department of Applied Mathematics and Theoretical Physics} \\
{\em University of Cambridge} \\
{\em Cambridge, CB3 0WA, UK} \\
\\
$^\ddag${\em Department of Mathematical Sciences}\\
{\em Loughborough University}\\
{\em Loughborough, Leics, LE11 3TU, UK} }
\date{}
\maketitle

\begin{abstract}
The problem of determining the neuronal current inside the brain
from measurements of the induced magnetic field outside the head
is discussed under the assumption that the space occupied by the
brain is approximately spherical. By inverting the Geselowitz
equation, the part of the current which can be reconstructed from
the measurements is precisely determined. This actually consists
of only certain moments of one of the two functions specifying the
tangential part of the current. The other function specifying the
tangential part of the current as well as the radial part of the
current are completely arbitrary. However, it is also shown that
with the assumption of energy minimization, the current can be
reconstructed uniquely. A numerical implementation of this unique
reconstruction is also presented.
\end{abstract}

\section{Introduction}

Magnetoencephalography (MEG) is a non invasive technique that can
be used to investigate brain activity. The physiological basis of
MEG is the following: The main functional units of the brain are
certain highly specialized cells called neurons. For higher mental
processes the most important part of the brain is its outermost
layer called cerebral cortex, which contains at least $10^{10}$
neurons. When neurons are active they produce small currents whose
basis is the change in the concentration of certain ions
\cite{hille} (ionic currents). The flow of current in the neural
system produces a weak magnetic field. The measurement of this
field outside the brain and the estimation of the current density
distribution that produced this field is called MEG. Other names
often used are magnetic source imaging, magnetic field tomography,
and current--flow imaging.

Neuromagnetic signals are typically 50--500$fT$, which are of the
order of $10^{-9}$ of the earth's geomagnetic field. Currently,
the only detector that can measure these tiny fields is the
superconducting quantum interference device (SQUID). The theory
and practice of SQUID as applied to MEG measurements, as well as
several practical approaches for shielding all other external
magnetic fields except that of the brain, are discussed in the
excellent review \cite{hhikl}. Here we only note that the SQUID,
which is the most sensitive detector of any kind available to
scientists \cite{clark}, is based on the exploitation of several
quantum--mechanical effects, including superconductivity as well
as the Josephson effect. The SQUID can be thought of as a digital
magnetometer where each ``digit'' represents one flux quantum, and
it is essentially a transducer converting a tiny change in
magnetic flux into a voltage. Whole--head magnetometer systems are
now used by several laboratories in Europe, USA and Japan.

The current density $\bf J$ and the magnetic field $\bf B$ are
related by the Maxwell equations. These equations can be
simplified using two facts. First, the permeability of the tissue
in the head denoted by $\mu$ is that of the free space, i.e.\ $\mu =
\mu_0$.  Second, the quasistatic approximation is valid, namely
the terms $\partial {\bf E} / \partial t$ and
$\partial {\bf B}/ \partial t$
can be neglected, where ${\bf E}$ denotes the electric field and
${\bf B}$ denotes the magnetic
induction\footnote{Let $\sigma$ and $\varepsilon$ denote conductivity
and permitivity which are assumed to be uniform, and let ${\bf E} =
{\bf E}_0({\bf x}) \exp (2 \pi i ft)$, where $f$ denotes frequency.
Then Maxwell equations imply that the term $\partial {\bf E} /
\partial t$ can be neglected provided that $|\varepsilon \partial
{\bf E} /\partial t| \ll |\sigma {\bf E}|$, or $2 \pi f
\varepsilon/ \sigma \ll 1$. This is indeed the case since for the
brain $\sigma = 0.3 \Omega^{-1} m^{-1}$, $\varepsilon=10^5
\varepsilon_0$, and since in neuromagnetism one usually deals with
frequencies of about $100Hz$ \cite{hhikl}, $2 \pi f
\varepsilon/\sigma \sim 2 \times 10^{-3}$. Similar arguments hold
true for the {\bf B} field.}.
Using these facts the Maxwell equations become
\be
\nabla \cdot {\bf B}=0, \quad \nabla \wedge {\bf B} = \mu_0 {\bf J},
\label{max1}
\ee
where $\cdot$ and $\wedge$ denote the scalar and vector product,
respectively, and $\nabla$ denotes the usual gradient. Part of
${\bf J}$ is due to neuronal activity, and part of ${\bf J}$ is
due to the electric field ${\bf E}$,
\be
{\bf J} = {\bf J^p} + \sigma {\bf E},
\label{max2}
\ee
where ${\bf J^p}$ denotes the neuronal current (primary current) and $\sigma$
denotes the conductivity. The electric field ${\bf E}$ satisfies $\nabla
\wedge {\bf E} = {\bf 0}$, thus there exists a scalar function $V$,
called the voltage potential, such that
\be
{\bf E} = -\nabla V.
\label{max3}
\ee
Making the further assumption that $\sigma = \sigma_I$ inside the head and
$\sigma = \sigma_O$ outside the head, where $\sigma_O$ and $\sigma_I$ are
constants, equations (\ref{max1})--(\ref{max3}) imply the celebrated
Geselowitz equation \cite{gese}
\begin{eqnarray}
{\bf B} ({\bf x}) & = & \frac{\mu_0}{4\pi} \int_{\Omega} {\bf J^p} ({\bf y})
\wedge \frac{{\bf x}-{\bf y}}{|{\bf x}-{\bf y}|^3} d {\bf y} \label{gese1} \\
& - & \frac{\mu_0}{4\pi}(\sigma_I-\sigma_O) \int_{\partial
\Omega} V({\bf y}) {\bf n}({\bf y}) \wedge
\frac{{\bf x}-{\bf y}}{|{\bf x}-{\bf y}|^3} dS, \quad {\bf x} \notin \Omega,
\nonumber
\end{eqnarray}
where $|{\bf x}|$ denotes the length of the vector ${\bf x}$,
$\Omega$ denotes the volume occupied by the head, $\partial\Omega$
is the boundary of $\Omega$, ${\bf n}$ denotes the unit outward
vector normal to the surface $\partial\Omega$, and $dS$ denotes
the infinitesimal surface element on $\partial \Omega$. For a recent
rigorous derivation of this equation see \cite{dakar}.

Equation (\ref{gese1}) relates ${\bf J^p}$ inside the head with
${\bf B}$ outside the head. However, it also involves the value of
$V$ on the surface of the head. This serious complication can be
avoided if one makes the simplifying assumption that the head is
spherical. Then, and if in addition $\sigma_O = 0$, which is
justified since $\sigma_O \ll \sigma_I$, equation (\ref{gese1})
reduces to \cite{grge}--\cite{sarvas}
\be
\left. \begin{array}{l}
 {\bf B} = \mu_0 \nabla U, \vspace{0.1cm} \\
 U({\bf x}) = {\displaystyle
 \frac{1}{4\pi} \int_{|{\bf y}| \le 1} \frac{{\bf J^p}({\bf y}) \wedge
 {\bf y} d{\bf y}}{|{\bf x}-{\bf y}| (|{\bf x}|
 |{\bf x}-{\bf y}| + {\bf x} \cdot ({\bf x}-{\bf y}))} \cdot {\bf x}}, \quad
 |{\bf x}| > 1.
       \end{array} \right.
\label{gese2}
\ee
Equation (\ref{gese2}) relates ${\bf J^p}$ inside the head $(|{\bf
x}|<1)$ with ${\bf B}$ outside the head. This equation is the
starting point of many of the algorithms used in MEG. It defines
the following inverse problem: Given ${\bf B}$, which is obtained
from the measurements, find ${\bf J^p}$.

The main difficulty with the above inverse problem is that it is
{\bf not} unique. This fact was already known to Helmholtz since
1853 \cite{helm}. For example, it is clear from equation
(\ref{gese2}) that the radial part of ${\bf J^p}$ does not
contribute to $U$. However, in spite of intense scrutiny by many
investigators, the fundamental question of {\it which part of
${\bf J^p}$ can be reconstructed} remained open.

Here we first give a complete answer to this question, see theorem
2: ${\bf J^p}$ can be uniquely decomposed in the form
\be
J^\rho {\bf e}_\rho +\frac{1}{\rho} \left( \frac{\partial
G}{\partial \theta} - \frac{1}{\sin \theta} \frac{\partial
F}{\partial \varphi} \right) {\bf e}_\theta + \frac{1}{\rho}
\left( \frac{1}{\sin \theta} \frac{\partial G}{\partial \varphi} +
\frac{\partial F}{\partial \theta} \right) {\bf e}_\varphi,
\label{unique}
\ee
where ${\bf e}_\rho$, ${\bf e}_\theta$, ${\bf e}_\varphi$ are the
unit vectors associated with the spherical coordinates $(\rho,
\theta, \varphi)$ and $J^\rho$, $G$, $F$ are scalar functions of
$(\rho, \theta, \varphi)$. This decomposition for vector fields on
the sphere is the analogue of the celebrated Helmholtz
decomposition for vector fields on $\mathbb{R}^3$. We will show
that knowledge of $U$ determines only certain moments of $F$ with
respect to $\rho$, while $J^\rho$ and $G$ are arbitrary. More
precisely, it can be shown that $U$ can be represented in the form
$U = \sum_{\ell,m} c_{\ell,m} \rho^{-(\ell + 1)} Y_{\ell, m}
(\theta, \varphi)$, where $Y_{\ell, m}$ are the usual spherical
harmonics and the constants $c_{\ell, m}$ are determined from the
measurements. Then we will show that $F$ can be represented in the
form $F = \sum_{\ell, m} f_{\ell, m}(\rho) Y_{\ell, m} (\theta,
\varphi)$, where only the moments of $f_{\ell, m}$ are determined
in terms of $c_{\ell, m}$,
\[ \ell \int^1_0 \rho^{\ell+1} f_{\ell, m} (\rho) d\rho =
(2\ell+1) c_{\ell,m}. \]

The above results imply that by decomposing ${\bf J^p}$ into a
``silent'' component and into an ``effective'' component, we can
show that the Geselowitz integral operator provides an one to one
map of the effective component of ${\bf J^p}$ into the magnetic
field {\bf B}, or into the magnetic potential $U$, outside the brain.
Furthermore, given $U$ the effective component can be explicitly
computed. We emphasise that, since the decomposition into a silent
and into an effective part is of a general nature independent of
any assumptions on ${\bf J^p}$, our result that $U$ determines the
effective component of the current uniquely and says nothing about
the silent component, is actually a general statement which is
{\it model independent}.

The next part of the paper deals with the case when we assume some
relations between the effective and the silent components: We will
show that if one requires that ${\bf J^p}$ is such that energy is
minimized, then ${\bf J^p}$ is indeed unique, see theorem 3: In
this case $J^\rho$, $G$, $F$ are given by the equations \be
J^\rho=G=0, \quad F = \sum_{\ell=1}^\infty \sum_{m=-\ell}^{\ell}
\frac{(2\ell+1)(2\ell+3)}{\ell}c_{\ell,m}
\rho^{\ell+1}Y_{\ell,m}(\theta,\varphi). \label{jgf} \ee

In addition to the above analytical results we also present a
numerical algorithm which given $U(\rho,\theta,\varphi)$ for one
specific value of $\rho>1$ and for some equally spaced values
$\{\theta_i\}_0^{i_{max}}$ and $\{\varphi_j\}_0^{j_{max}}$, it
first computes $c_{\ell,m}$ and then computes ${\bf J^p}$ using
equations (\ref{unique}) and (\ref{jgf}).

The non uniqueness of the inverse problem has been the Achilles
heel of MEG. For example in the most comprehensive review on MEG
\cite{hhikl}, it is written that ``with the assumption that MEG
mainly reflects the activity in the tangential part of the
cortical currents'', while in \cite{ibc} it is written ``what
cannot be seen should not be looked for''. Even the ``father'' of
MEG, D.\ Cohen has stated \cite{see} ``identifying those
tangential sources, rather than localization, is the real use of
the MEG, there is no localization magic''. We hope that both the
analytical and the numerical results presented here will
contribute towards determining the advantages as well as the
limitations of MEG.

Regarding other brain imaging techniques we note that at present
time the most important such techniques are the functional
magnetic resonance imaging (fMRI) the positron emission tomography
(PET) and the single photon emission computed tomography (SPECT),
as well as a new version of electroencephalography (EEG). These
techniques involve tradeoffs among the following important
considerations: temporal resolution, spatial resolution,
invasiveness, and cost. Assuming that the question of uniqueness
of the MEG is answered, the spatial resolution of MEG ($1 cm$), of
PET and SPECT (4--5$mm$), and of fMRI ($1.5 mm$) are similar; the
spatial resolution of the conventional EEG is quite poor. On the
other hand the time resolution of EEG and MEG is much better than
that of PET, SPECT and fMRI. The time resolution of PET, SPECT and
MRI is of the order of 1 second, while that of MEG and EEG is of
the order of 10 milliseconds. This is a crucial factor if one
wants to study brain dynamics. For example, MEG data suggest that
speech areas of the brain are activated 100 milliseconds after the
visual areas. MEG is the only truly non invasive method. EEG is
minimally invasive (placing electrodes on the scalp), while in
PET, SPECT and MRI the subject is exposed to radioactive tracers
and to strong magnetic fields, respectively. EEG requires a rather
inexpensive apparatus (of the order of thousands of dollars). The
fMRI has the advantage that can be obtained by modifying the
existing MRI apparatus. PET employs positron--emitting
radionuclides which have such short half--lives that there must be
a cyclotron near the site of scanning, thus the cost is of the
order of multimillion dollars. The cost of the MEG is similar to
that of the PET.

We conclude the introduction with some remarks:
\begin{itemize}
\item[(a)] We expect that the combination of our analysis of the spherical
model and perturbation theory can be used to study realistic head
geometries. In this respect we also note that progress has been
recently made regarding ellipsoidal geometry \cite{daka}.
\item[(b)] The question of what additional information can one obtain by
measuring ${\bf E}$ (using EEG) is under investigation.
\item[(c)] Due to the orthogonality of the decomposition of ${\bf J^p}$
into silent and effective components, the assumptions that the $L^2$ norm
of the solution is minimal, implies that the silent component vanishes.
Clearly, one can assume other relations between the silent and the effective
components, for example one may assume that the current consists of a finite
number of dipoles. It is well known that this assumption, under certain
conditions, also leads to a unique solution. This current can also be
represented in the form (\ref{unique}) with $F$ of a particular form,
and therefore can be considered within our formulation. Thus the answer
becomes model dependent only at the stage when one makes an assumption about
the form of ${\bf J^p}$. For other models see \cite{sch}-\cite{hail}.
\item[(d)] ``Least--square'' methods have been used extensively in
inverse problems. However, our approach of using such methods in order to
find an approximate numerical solution of the Geselowitz equation is
fundamentally different than the existing ones. Indeed, it is based on
the explicit decomposition of the current into a silent and an effective
component, and thus could not have been used before obtaining this
decomposition.
\item[(e)] In practice, the magnetic field is measured approximately over
a half--sphere over the head and {\it not} over a whole sphere. However,
since in the numerical reconstruction we assume a {\it finite} number of
spherical harmonics, the approximate knowledge of $U$ over part of a sphere
is sufficient to determine approximately the current. Clearly the problem
becomes more and more ill-posed when the number of spherical harmonics
increases. A stability result in this direction is under investigation.
\item[(f)] It has been correctly pointed out by one of the referees that
it is sufficient for the solution of the inverse problem to invert
$\partial U/\partial_{|{\bf x}|}$ instead of $U$. Furthermore it has been
correctly pointed out that this latter inversion is much simpler since
the expression for $\partial U/\partial_{|{\bf x}|}$ is simpler than
the expression of $U$ (see \cite{dakar}).
\item[(g)] A short summary of the analytical results
presented here was announced in \cite{fgk}.
\end{itemize}

\section{Analytical Results}
\setcounter{equation}{0}

We first show that equations (\ref{gese2}) can be written in an
alternative form, which is more convenient for determining the
part of ${\bf J^p}$ which can be reconstructed from the knowledge
of $U({\bf x})$.

\vskip 0.3cm
\noindent
{\bf Theorem 1.}
Let $U({\bf x})$ be defined in terms of ${\bf J^p}$ by equation
(\ref{gese2}). Then $U({\bf x})$ can also be expressed by the
alternative representation
\be
U({\bf x}) = -\frac{1}{4\pi} \int_{|{\bf y}| \le 1}
\frac{1}{|{\bf x}-{\bf y}|} \left( \frac{1}{|{\bf y}|^2}
\int^1_{|{\bf y}|} \{ (\nabla_{\bf z} \wedge {\bf J^p}({\bf z}))
\cdot {\bf z} \}_{{\bf z}=\frac{|{\bf z}|}{|{\bf y}|}
{\bf y}} |{\bf z}| d|{\bf z}| \right) d{\bf y}, \quad |{\bf x}|>1.
\label{theo1}
\ee

\vskip 0.3cm
\noindent
{\bf Proof.}
Let $I({\bf z})$ denote the following function of ${\bf z}$,
\be
I({\bf z}) = \frac{4\pi}{|{\bf z}|} \int^{|{\bf z}|}_0
\{ ({\bf J^p}({\bf z}) \wedge {\bf x}) \cdot (\nabla_{\bf x}
\Phi({\bf x})) \}_{{\bf x}=\frac{|{\bf x}|}{|{\bf z}|}
{\bf z}} d|{\bf x}|,
\label{relI}
\ee
where $\Phi({\bf x}) \in C^\infty_0 (\mathbb{R}^3)$. We will
integrate $I({\bf z})$ over the sphere $|{\bf z}| \le 1$: We first
multiply by $|{\bf z}|^2$ and integrate with respect to $d|{\bf
z}|$ along $0<|{\bf z}|<1$. Interchanging in the resulting
expression the order of the integration with respect to $d|{\bf
x}|$ and to $d|{\bf z}|$ we find
\[ \int_0^1 I({\bf z}) |{\bf z}|^2 d|{\bf z}| = 4\pi \int^1_0
\{ \nabla_{\bf x} \Phi({\bf x}) \}_{{\bf x}=\frac{|{\bf x}|}{|{\bf
z}|} {\bf z}} \cdot \left( \int^1_{|{\bf x}|}
{\bf J^p}({\bf z}) \wedge |{\bf x}| {\bf z} \, d|{\bf z}| \right)
d|{\bf x}|. \]
We then integrate this equation with respect to $d {\bf {\hat z}}$,
${\bf {\hat z}}={\bf z}/|{\bf z}|$, and denote $|{\bf x}|{\bf {\hat z}}$
by ${\bf y}$. This yields
\be
\int_{|{\bf z}| \le 1} I({\bf z}) d{\bf z} =
-4 \pi \int_{|{\bf y}| \le 1} \Phi({\bf y}) \nabla_{{\bf y}} \cdot
\left( \frac{1}{|{\bf y}|^2} \int^1_{|{\bf y}|} {\bf J^p}
\left(\frac{|{\bf z}|}{|{\bf y}|} {\bf y}\right) \wedge {\bf y} |{\bf z}|
\, d|{\bf z}| \right) d{\bf y}.
\label{rely}
\ee
It is straightforward to show that
\be
\nabla_{{\bf y}} \cdot \left( {\bf J^p} \left(\frac{|{\bf z}|}{|{\bf y}|}
{\bf y}\right) \wedge {\bf y} \right) = \frac{|{\bf z}|}{|{\bf y}|}
\{ \nabla_{{\bf z}} \wedge {\bf J^p} ({\bf z})
\}_{{\bf z}=\frac{|{\bf z}|}{|{\bf y}|} {\bf y}} \cdot {\bf y}.
\label{relyn}
\ee
Indeed, the rhs of this equation equals
\be
\frac{|{\bf z}|}{|{\bf y}|} \left( y_2 \frac{\partial J^p_1}{\partial
z_3}-y_1 \frac{\partial J^p_2}{\partial z_3} \right) +cp,
\label{part}
\ee
where $cp$ denotes cyclic permutation; the lhs of equation
(\ref{relyn}) equals
\[ \frac{\partial}{\partial y_3} (y_2 J^p_1 - y_1 J^p_2) + cp, \]
and using the chain rule as well as noting that several of the resulting
terms cancel we find the expression (\ref{part}).

Using equation (\ref{relyn}), as well as noting that the term
$\nabla_{\bf y}(|{\bf y}|^{-2})$ is perpendicular to ${\bf J^p} \wedge
{\bf y}$, the rhs of equation (\ref{rely}) becomes
\[ -4\pi \int_{|{\bf y}| \le 1} \Phi({\bf y}) \frac{1}{|{\bf y}|^2}
\left( \int^1_{|{\bf y}|} \{ (\nabla_{\bf z} \wedge {\bf J^p}({\bf z}))
\cdot {\bf z} \}_{{\bf z}=\frac{|{\bf z}|}{|{\bf y}|} {\bf y}}
|{\bf z}| d|{\bf z}| \right) d{\bf y}. \]
Replacing the rhs of equation (\ref{rely}) by this expression and
replacing $I({\bf z})$ by the definition (\ref{relI}), equation
(\ref{rely}) and the standard Green's function representation for
solutions of Poisson's equation, give equation (\ref{theo1})
provided that the result of the lemma proven in the appendix A is
valid. Note that according to our proof, equation (\ref{theo1}) is
valid in the distributional sense, but simple regularity arguments
imply that it is also valid pointwise. \hfill {\bf QED}

\vskip 0.3cm
\noindent
{\bf Theorem 2 (Representation theorem)}
\newline
\noindent
The vector ${\bf J^p}({\bf x})$ can be uniquely decomposed in the
form
\be
{\bf J^p}({\bf x}) = J^\rho (\rho,\theta,\varphi)
{\bf e}_\rho + J^\theta (\rho,\theta,\varphi) {\bf e}_\theta +
J^\varphi (\rho,\theta,\varphi) {\bf e}_\varphi,
\label{jp}
\ee
where ${\bf e}_\rho$, ${\bf e}_\theta$, ${\bf e}_\varphi$ are the
unit vectors associated with the spherical coordinates $\rho>0$,
$0 \le \theta \le \pi$, $0 \le \varphi < 2\pi$, and the scalar
functions $J^\theta$ and $J^\varphi$ can be represented in the form
\be
J^\theta = \frac{1}{\rho} \left( \frac{\partial G}{\partial
\theta} - \frac{1}{\sin \theta} \frac{\partial F}{\partial \varphi}
\right), \quad J^\varphi = \frac{1}{\rho} \left( \frac{1}{\sin \theta}
\frac{\partial G}{\partial \varphi} + \frac{\partial F}{\partial \theta}
\right),
\label{jthph}
\ee
where $G(\rho,\theta,\varphi)$ and $F(\rho,\theta,\varphi)$ are scalar
functions of the arguments included.

Assume that $U({\bf x})$ is defined in terms of ${\bf J^p}$
by equation (\ref{gese2}). Then
\be
U({\bf x}) = -\frac{1}{4 \pi}
\int_{|{\bf y}| \le 1} \frac{1}{|{\bf x}-{\bf y}|} \left(
\frac{1}{|{\bf y}|^2} \int^1_{|{\bf y}|}
\Delta_{\theta,\varphi}F(|{\bf z}|,\theta,\varphi)
d|{\bf z}| \right) d{\bf y}, \quad |{\bf x}|>1,
\label{gese3}
\ee
where $\Delta_{\theta,\varphi}$ denotes the Laplacian with respect
to the spherical coordinates $\theta$ and $\varphi$, i.e.\
\[ \Delta_{\theta,\varphi} = \frac{1}{\sin \theta} \left[
\frac{\partial}{\partial \theta} \left( \sin \theta
\frac{\partial}{\partial \theta} \right) +
\frac{1}{\sin \theta} \frac{\partial^2}{\partial \varphi^2}
\right]. \]

\vskip 0.3cm
\noindent
{\bf Proof.}
We first decompose ${\bf J^p}$ into a radial and a tangential
component. Clearly $J^\rho$ gives no contribution to $U$. Also the
tangential component can be {\it uniquely} decomposed in the form
(\ref{jthph}), see appendix B. Using equations (\ref{jp}) and
(\ref{jthph}) we find
\[ (\nabla \wedge {\bf J^p}) \cdot {\bf z} =
\frac{1}{|{\bf z}|} \left( \frac{1}{\sin\theta}
\frac{\partial}{\partial \theta} \sin \theta \frac{\partial
F}{\partial \theta} + \frac{1}{\sin^2 \theta} \frac{\partial^2 F}{\partial
\varphi^2}\right), \]
and (\ref{theo1}) becomes equation (\ref{gese3}). \hfill {\bf QED}

\vskip 0.3cm
\noindent
{\bf Corollary (Non uniqueness of the inverse problem)}
\newline
\noindent
Assume that $U({\bf x})$ is defined in terms of ${\bf J^p}$
by equation (\ref{gese2}). Let a vector ${\bf J^p}({\bf x})$ be written
in the form (\ref{jp}) where the scalar functions $J^\theta$ and
$J^\varphi$ are given in terms of the scalar function
$G$ and $F$ by equation (\ref{jthph}).

The function $U({\bf x})$ is {\it independent} of $J^\rho$ and of $G$,
and furthermore only certain moments of $F$ can be computed in
terms of $U$. In particular,
$F(\rho,\theta,\varphi)$ is given by the expression
\[ F(\rho,\theta,\varphi) = \sum_{\ell=1}^\infty
\sum_{m=-\ell}^{\ell} f_{\ell, m}(\rho) Y_{\ell, m} (\theta,
\varphi), \quad \rho<1, \quad
0 \le \theta \le \pi, \quad 0 \le \varphi < 2\pi, \]
where $Y_{\ell, m}$ are the usual spherical harmonics,
the moments of $f_{\ell, m}(\rho)$ can be determined in terms of
$c_{\ell, m}$,
\be
\ell \int^1_0 \rho^{\ell+1} f_{\ell,m}(\rho) d\rho =
(2\ell+1) c_{\ell,m},
\label{rell}
\ee
and the constants $c_{\ell, m}$ can be determined from the given
data using the fact that $U({\bf x})$ can be expressed in the form
\be
U(\rho,\theta,\varphi) = \sum_{\ell=1}^\infty
\sum_{m=-\ell}^{\ell} c_{\ell, m} \rho^{-(\ell + 1)}
Y_{\ell, m}(\theta,\varphi), \quad \rho>1, \quad
0 \le \theta \le \pi, \quad 0 \le \varphi \le 2\pi.
\label{relU}
\ee

\vskip 0.3cm
\noindent
{\bf Proof.}
Equation (\ref{gese3}) implies
\be
\left. \begin{array}{l}
 {\displaystyle \Delta U = \frac{1}{|{\bf x}|^2} \int^1_{|{\bf x}|}
 \Delta_{\theta,\varphi} F
 (|{\bf z}|,\theta,\varphi) d|{\bf z}|, \quad
 |{\bf x}|<1,} \vspace{0.1cm} \\
 \Delta U = 0, \quad |{\bf x}|>1.
       \end{array} \right.
\label{deltaU}
\ee
Let us represent $F$ and $U$ in terms of spherical harmonics by
\[ F(\rho,\theta,\varphi) = \sum_{\ell, m} f_{\ell,m}(\rho) Y_{\ell,m}
(\theta,\varphi)\ \quad\mbox{and}\quad U(\rho,\theta,\varphi) = \sum_{\ell, m}
u_{\ell,m}(\rho) Y_{\ell,m} (\theta,\varphi). \]
Then equations (\ref{deltaU}) imply
\[ u''_{\ell, m} + \frac{2}{\rho} u'_{\ell, m} - \frac{\ell(\ell +
1)}{\rho^2} u_{\ell, m} = \left\{ \begin{array}{cl}
-\frac{\ell(\ell + 1)}{\rho^2} \int^1_\rho f_{\ell, m} (\rho') d\rho'
& \rho<1 \\
0 & \rho>1,
\end{array} \right. \]
where prime denotes differentiation with respect to $\rho$. The
general solution of the homogeneous problem is $\alpha\rho^\ell +
\beta\rho^{-(\ell + 1)}$, where $\alpha$ and $\beta$ are constants
and $\ell$ is a positive integer. Since $u_{\ell, m} \to 0$ as
$\rho\to\infty$ it follows that
\[ u_{\ell, m} = c_{\ell, m} \rho^{-(\ell + 1)}. \]
To solve the inhomogeneous problem we use variation of parameters
in the form $u_{\ell, m}(\rho) = A_{\ell, m} (\rho)\rho^\ell$.
This implies
\[ (A'_{\ell, m} \rho^{2\ell + 2})' = -\ell(\ell + 1) \alpha_{\ell,
m}(\rho), \quad \alpha_{\ell,m}(\rho) \doteqdot \rho^\ell \int^1_\rho
f_{\ell,m}(\rho') d\rho'. \]
Thus
\[ A'_{\ell, m} \rho^{2\ell + 2} = \ell(\ell + 1) \int^1_\rho \alpha_{\ell,
m}(\rho') d\rho' + A'_{\ell, m}(1). \]

Convergence at $\rho = 0$ implies
\be
A'_{\ell, m}(1) + \ell(\ell + 1) \int^1_0
\alpha_{\ell, m}(\rho') d\rho' = 0.
\label{Almp}
\ee
Using $A_{\ell, m} = u_{\ell, m} \rho^{-\ell}$, we find
\[ A'_{\ell, m}(1) = u'_{\ell, m}(1) - \ell u_{\ell, m}(1) =
\left(c_{\ell, m} \rho^{-(\ell + 1)}\right)'\bigg|_{\rho = 1}
-\ell c_{\ell, m} \rho^{-(\ell + 1)}\bigg|_{\rho = 1} =
-(2\ell + 1) c_{\ell, m}. \]
This equation together with (\ref{Almp}) imply
\[ \ell(\ell + 1) \int^1_0 \alpha_{\ell, m} (\rho) d\rho =
(2\ell + 1) c_{\ell, m}. \]
Using integration by parts we find (\ref{rell}). \hfill {\bf QED}

\vskip 0.3cm
\noindent
{\bf Theorem 3 (Minimization of energy)}
\newline
\noindent
Define the energy by
\be
W \doteqdot \int_{|{\bf x}| \le 1} |{\bf J^p}|^2 d{\bf x}.
\label{energy}
\ee
Then if
\[ {\bf J^p} = J^\rho {\bf e}_\rho + J^\theta {\bf e}_\theta +
J^\varphi {\bf e}_\varphi, \]
where $J^\theta$ and $J^\varphi$ are given by equations
(\ref{jthph}), it follows that the minimum of $W$ under the constrain
\[ F = \sum_{\ell=1}^\infty \sum_{m=-\ell}^{\ell}
f_{\ell,m}(\rho) Y_{\ell,m}(\theta,\varphi),
\quad \ell \int^1_0 \rho^{\ell+1} f_{\ell,m}(\rho) d\rho =
(2\ell+1)c_{\ell,m}, \]
where $Y_{\ell,m}$ are the usual spherical harmonics and
$c_{\ell,m}$ are given constants, is achieved when
\be
J^\rho = G=0, \quad F = \sum_{\ell=1}^\infty \sum_{m=-\ell}^{\ell}
\frac{(2\ell+1)(2\ell+3)}{\ell} c_{\ell,m} \rho^{\ell+1}
Y_{\ell,m}(\theta,\varphi).
\label{jrgf}
\ee

\vskip 0.3cm
\noindent
{\bf Proof.}
Substituting equation (\ref{jp}) and (\ref{jthph}) in the rhs of
equation (\ref{energy}) we find
\[ W = \int_{|{\bf x}| \le 1} \!\left[ (J^\rho)^2 + \frac{1}{\rho^2} \left(
\frac{\partial G}{\partial \theta} \right)^2 + \frac{1}{\rho^2\sin^2\theta}
\left( \frac{\partial G}{\partial\varphi} \right)^2 +
\frac{1}{\rho^2\sin^2\theta}
\left( \frac{\partial F}{\partial\varphi} \right)^2 + \frac{1}{\rho^2}
\left( \frac{\partial F}{\partial \theta} \right)^2 \right] \! d{\bf x}, \]
where we have used that the term involving
$G_\varphi F_\theta - G_\theta F_\varphi$ vanishes,
\[ \int^1_0 \int^\pi_0 \int^{2\pi}_0
\frac{1}{\rho^2\sin\theta} \left[ -\frac{\partial G}{\partial \theta}
\frac{\partial F}{\partial\varphi} + \frac{\partial G}{\partial\varphi}
\frac{\partial F}{\partial \theta} \right]
\rho^2\sin\theta d\rho d\theta d\varphi = 0. \]

The constraint involves only $F$, thus it follows that the minimal energy
is achieved when $J^\rho=G=0$ and when $H$ is minimal, where
\be
H = \int^1_0 \int^\pi_0 \int^{2\pi}_0 \left[
\frac{1}{\rho^2\sin^2\theta} \left( \frac{\partial F}{\partial
\varphi} \right)^2 + \frac{1}{\rho^2} \left(
\frac{\partial F}{\partial \theta} \right)^2 \right]
\rho^2\sin\theta d\rho d\theta d\varphi.
\label{Hcap}
\ee
The term inside the bracket equals $|\nabla F|^2 - (\frac{\partial
F}{\partial \rho})^2$, which using integration by parts (with either
$F$ or $\frac{\partial F}{\partial\rho}$ equal to 0 at $|{\bf x}|=1$),
equals $-[F\Delta F +(\frac{\partial F}{\partial \rho})^2]$, where
\[ \Delta F = \frac{\partial^2F}{\partial\rho^2} + \frac{2}{\rho}
\frac{\partial F}{\partial \rho} + \frac{1}{\rho^2}
\Delta_{\theta,\varphi}F. \]
Using
\[ F=\sum_{\ell,m} f_{\ell,m}(\rho) Y_{\ell,m}(\theta,\varphi), \quad
\Delta_{\theta,\varphi} Y_{\ell,m} = -\ell(\ell+1) Y_{\ell,m}, \]
and the orthogonality of the spherical harmonics, it follows that
\[ H = -\sum_{\ell,m} \int^1_0 \left\{ \left[ f''_{\ell,m}(\rho) +
\frac{2}{\rho} f'_{\ell,m}(\rho) - \frac{\ell(\ell+1)}{\rho^2}
f_{\ell,m}(\rho) \right] f_{\ell,m}(\rho) + (f'_{\ell,m}(\rho))^2
\right\} \rho^2 d\rho. \]
Hence,
\[ H = -\sum_{\ell,m} \left[ \int^1_0 \left\{ (f_{\ell,m}
f'_{\ell,m}\rho^2)' - \ell(\ell+1) f^2_{\ell,m}(\rho) \right\}
d\rho \right]. \]
Thus, provided that either $f_{\ell,m}(1)$ or $f'_{\ell,m}(1)$
equals zero\footnote{These conditions are true since the
support of ${\bf J^p}$ lies in the interior of the sphere.},
we find
\[ H = \sum_{\ell,m} \ell(\ell+1) \int^1_0 f^2_{\ell,m}(\rho) d\rho. \]
The assumption that $f_{\ell, m}(1) = 0$ is without loss of
generality since the tangential part of the energy which is given
by equation (\ref{Hcap}) does not involve differentiation over
$\rho$, thus in general (\ref{Hcap}) can be obtained by
approximating $f$ by functions equal to zero at $\rho = 1$ and
then passing to the limit.

The minimization of this $H$, under the constraint (\ref{rell}),
implies (\ref{jrgf}).

We note that equation (\ref{gese2}) implies that $U({\bf x})$
behaves like $0(\rho^{-2})$, hence $\ell > 0$ in equation
(\ref{relU}), $c_{00} = 0$, and the sum (\ref{jrgf}) starts with
$\ell = 1$. \hfill {\bf QED}

\section{Numerical Implementation}
\setcounter{equation}{0}

In equation (\ref{relU}) $Y_{\ell,m}$ denotes the
spherical harmonics, namely
\be
\left. \begin{array}{l}
Y_{\ell,m}(\theta,\varphi)=a_{\ell,m} P_{\ell,m}(\cos \theta) e^{im \varphi},
\vspace{0.1cm} \\
Y_{\ell,-m} = (-1)^m \overline{Y_{\ell,m}}, \qquad \ell \ge 1,
\quad 0 \le m \le \ell,
       \end{array} \right.
\label{ylm}
\ee
where the bar denotes complex conjugate and
\be
a_{\ell,m} =\sqrt{ \frac{2\ell+1}{4 \pi} \frac{(\ell-m)!}{(\ell+m)!} }.
\label{alm}
\ee
$P_{\ell,m}$ are the Legendre functions, namely
\[ P_{\ell,m}(x) = (-1)^m (1-x^2)^{m/2} \frac{d^m}{dx^m} P_\ell(x), \]
where
\[ P_\ell(x) = \frac{1}{2^\ell \ell!}\frac{d^\ell}{dx^\ell}(x^2-1)^\ell \]
are the usual Legendre polynomials of degree $\ell$.

For the numerical implementation we replace in the sums appearing
in (\ref{relU}), (\ref{jrgf}) $\infty$ by $\ell_{max}$, where
$\ell_{max}$ is chosen by the procedure explained below.

\subsection{Computation of $c_{\ell,m}$}

We first discuss how to compute $c_{\ell,m}$ from either
$U(\rho,\theta,\varphi)$ or from ${\bf B}(\rho,\theta,\varphi)$.

Suppose we know $U(\rho,\theta,\varphi)$ for one specific value of
$\rho>1$ and for some equally spaced values
$\theta_i$, $\varphi_j$, such us
\begin{eqnarray*}
& & 0 \le \theta_i \le  \pi, \quad i=0,\ldots,i_{max}, \\
& & 0 \le \varphi_j < 2 \pi, \quad j=0,\ldots,j_{max}.
\end{eqnarray*}
Using the orthogonality of $Y_{\ell,m}$ equation (\ref{relU}) implies
\[ \int_0^{2 \pi} \left( \int_{-1}^1 U(\theta,\varphi)
\overline{Y_{\ell,m}(\theta,\varphi)} d(\cos \theta) \right) d\varphi =
c_{\ell,m} \rho^{-(\ell+1)}. \]
Therefore, using the first equation in (\ref{ylm}), we obtain
\[ c_{\ell,m} = \rho^{\ell+1} a_{\ell,m} \int_0^{2 \pi} \left( \int_0^\pi
U(\theta,\varphi) P_{\ell,m}(\cos \theta) \sin \theta d \theta \right) e^{-im
\varphi} d \varphi. \]
Using (\ref{ylm}), we find
\be
\left. \begin{array}{l}
c_{\ell,m} = \rho^{\ell+1} a_{\ell,m} {\hat U}_{\ell,m}, \vspace{0.1cm} \\
c_{\ell,-m} = (-1)^m \overline{c_{\ell,m}}, \qquad \ell \ge 1,
\quad 0 \le m \le \ell,
       \end{array} \right.
\label{clm}
\ee
where
\be
{\hat U}_{\ell,m} = \int_0^{2 \pi} {\tilde U}_{\ell,m}(\varphi) \cos m \varphi
d \varphi - i \int_0^{2 \pi} {\tilde U}_{\ell,m}(\varphi) \sin m \varphi
d \varphi,
\label{uhat}
\ee
and
\be
{\tilde U}_{\ell,m}(\varphi) = \int_0^{\pi} U(\theta,\varphi) P_{\ell,m}
(\cos \theta) \sin \theta d \theta.
\label{util}
\ee

For the numerical calculation of the three integrals appearing in (\ref{uhat})
and (\ref{util}) we use an extended closed formula, namely
\[ \int_{x_1}^{x_n} f(x)dx = \Delta x \left( \frac{3}{8} f_1 + \frac{7}{6} f_2
+ \frac{23}{24} f_3 + f_4 + \ldots + f_{n-3} + \frac{23}{24} f_{n-2} +
\frac{7}{6} f_{n-1} + \frac{3}{8} f_n \right). \]
For the numerical calculation of the Legendre functions
$P_{\ell,m}(\cos \theta)$
we use subroutine \texttt{plgndr} from Numerical Recipes \cite{recipes}.
The constants $a_{\ell,m}$ are given by (\ref{alm}).

Suppose we know ${\bf B}=(B_1,B_2,B_3)$ instead of $U$. Then, using the first
relation in (\ref{gese2}) and spherical coordinates we obtain
\be
U_\rho = \sin \theta \cos \varphi {\tilde B}_1 + \sin \theta \sin \varphi
{\tilde B}_2 + \cos \theta {\tilde B}_3,
\label{ur}
\ee
where
\[ {\tilde B}_i = \frac{B_i(\rho,\theta,\varphi)}{\mu_0}, \quad i=1,2,3. \]
Moreover, by differentiating (\ref{relU}) with respect to $\rho$ we find
\be
U_\rho(\rho,\theta,\varphi) =  - \sum_{\ell=1}^{\ell_{max}}
\sum_{m=-\ell}^\ell c_{\ell,m} (\ell+1) \rho^{-(\ell+2)}
Y_{\ell,m}(\theta,\varphi).
\label{ursum}
\ee
Thus, if we know ${\bf B}$, we can compute $U_\rho$ from (\ref{ur}) and then
we can compute $c_{\ell,m}$ from (\ref{ursum}), following the same procedure
as before.

{\bf The choice of $l_{max}$.} Using (\ref{ylm}) and the second relation in
(\ref{clm}), the real part of (\ref{relU}) implies
\begin{eqnarray}
\lefteqn{U(\rho,\theta,\varphi) = \sum_{\ell=1}^{\ell_{max}}
\rho^{-(\ell+1)} \cdot} \label{ufin} \\
& & \cdot \left(\! Re(c_{\ell,0}) a_{\ell,0} P_{\ell,0}(\cos
\theta)+2 \sum_{m=1}^\ell (Re(c_{\ell,m}) \cos m
\varphi-Im(c_{\ell,m}) \sin m \varphi) a_{\ell,m} P_{\ell,m}(\cos
\theta) \!\right). \nonumber
\end{eqnarray}
Differentiation with respect to $\rho$ yields
\begin{eqnarray}
\lefteqn{U_\rho(\rho,\theta,\varphi) = -\sum_{\ell=1}^{\ell_{max}}
(\ell+1) \rho^{-(\ell+2)} \cdot} \label{urfin} \\
& & \cdot \left(\! Re(c_{\ell,0}) a_{\ell,0} P_{\ell,0}(\cos
\theta)+2 \sum_{m=1}^\ell (Re(c_{\ell,m}) \cos m
\varphi-Im(c_{\ell,m}) \sin m \varphi) a_{\ell,m} P_{\ell,m}(\cos
\theta) \!\right). \nonumber
\end{eqnarray}

Therefore, after calculating the coefficients $c_{\ell,m}$
following the procedure outlined earlier we can use either
(\ref{ufin}) or (\ref{urfin}) to re--evaluate either $U$ or
$U_\rho$. In this way not only can we test the efficiency of our
procedure, but we can also run our program several times, in order
to find the most appropriate value for $\ell_{max}$.

\subsection{Computation of the Minimizing Current}

Using relations (\ref{jrgf}), (\ref{ylm}) and the second relation
in (\ref{clm}), the real parts of the functions $J^\theta$,
$J^\varphi$ defined in (\ref{jthph}) (with $G=0$) are given by
\begin{eqnarray}
\lefteqn{J^\theta(\rho,\theta,\varphi) = \frac{2}{\sin \theta}
\sum_{\ell=1}^{\ell_{max}} \frac{(2\ell+1)(2\ell+3)}{\ell} \rho^\ell
\cdot} \label{jth} \\
& & \cdot \left( \sum_{m=1}^\ell m (Re(c_{\ell,m}) \sin m\varphi+
Im(c_{\ell,m}) \cos m\varphi) a_{\ell,m} P_{\ell,m}(\cos \theta) \right),
\nonumber
\end{eqnarray}
and
\begin{eqnarray}
\lefteqn{J^\varphi(\rho,\theta,\varphi) = -\sin \theta
\sum_{\ell=1}^{\ell_{max}}
\frac{(2\ell+1)(2\ell+3)}{\ell} \rho^\ell \cdot} \label{jph} \\
& & \cdot \left(\! Re(c_{\ell,0}) a_{\ell,0} P'_{\ell,0}(\cos \theta)+2
\sum_{m=1}^\ell (Re(c_{\ell,m}) \cos m \varphi-Im(c_{\ell,m}) \sin m \varphi)
a_{\ell,m} P'_{\ell,m}(\cos \theta) \!\right). \nonumber
\end{eqnarray}
Recall that the Legendre functions satisfy the recurrence relation
\[ P'_{\ell,m}(x) = -\frac{mx}{1-x^2} P_{\ell,m}(x)-\frac{1}{\sqrt{1-x^2}}
P_{\ell,m+1}(x). \]
Therefore
\be
\left. \begin{array}{l}
-\sin \theta P'_{\ell,0}(\cos \theta) = P_{\ell,1}(\cos \theta),
\quad\mbox{for}\quad m=0, \vspace{0.1cm} \\
-\sin \theta P'_{\ell,m}(\cos \theta) = {\displaystyle \frac{m \cos
\theta}{\sin  \theta}} P_{\ell,m}(\cos \theta)+P_{\ell,m+1}(\cos \theta),
\quad\mbox{for}\quad m>0.
       \end{array} \right.
\label{plmp}
\ee

Thus, in order to calculate numerically the current we apply the
following procedure: We take some $\theta$ and $\varphi$ points,
such that $0 \le \theta \le \pi$, $0 \le \varphi \le 2\pi$. We
first calculate the Legendre functions $P_{\ell,m}(\cos \theta)$.
In a separate subroutine we calculate the quantities
$P_{\ell,m}(\cos \theta)/\sin \theta$ (for this purpose we have
developed a subroutine similar to \texttt{plgndr}). These
quantities appear in both (\ref{jth}) and the second relation of
(\ref{plmp}). Note that these quantities are valid  even for
$\theta=0$ or $\theta=\pi$. We then calculate from (\ref{plmp})
the quantities $-\sin \theta P'_{\ell,m}(\cos \theta)$. Finally,
we take a value of $\rho$ such as $0<\rho<1$ and calculate
$J^\theta(\rho,\theta,\varphi)$ from (\ref{jth}) and
$J^\varphi(\rho,\theta,\varphi)$ from (\ref{jph}). In all the
above we use the $\ell_{max}$ value that was found with the
procedure outlined in the previous subsection.

\subsection{Verification of the Algorithm}

We have tested our numerical algorithm for several functions
$U(\rho,\theta,\varphi)$. In what follows we discuss two
typical examples.

\vskip 0.3cm
\noindent
{\bf Example 1}
\newline
\noindent
Let $U$ be given by
\[ U(\rho,\theta,\varphi) = -2 \cos \theta \frac{1}{\rho^2}+\sin \theta
\cos \theta \cos \varphi \frac{1}{\rho^3}-\sin^2 \theta \cos 2\varphi
\frac{1}{\rho^3}. \]
Note that this function has the form (\ref{relU}) with $c_{\ell,m}=0$
for $\ell>2$.

First, we evaluate $U$ for $\rho=1.5$ and some equally spaced
$\theta_i$ and $\varphi_j$, where $i_{max}=100$, $j_{max}=200$. We
calculate numerically the coefficients $c_{\ell,m}$ from the first
relation of (\ref{clm}), (\ref{uhat}) and (\ref{util}), and then
evaluate from (\ref{ufin}) $U_a$, the approximate value of $U$, at the
above $\rho$, $\theta_i$ and $\varphi_j$. Furthermore, we start
with $U_\rho$ instead of a $U$, we calculate $c_{\ell,m}$ in a
similar way and then calculate the approximate value of $U_\rho$
from (\ref{urfin}).

We run our program several times with $l_{max}$ from 1 up to 40 and
we found that the best value is $l_{max}=2$, which is consistent with
the exact form of $U$. For this value the difference $|U-U_a|$ is of
order $10^{-7}$, at most.

Secondly, we calculate numerically $J^\theta$, $J^\varphi$, using
(\ref{jth})--(\ref{plmp}), in the above $\theta_i$, $\varphi_j$
and some equally spaced $\rho_k$, such as $0 \le \rho_k \le 1$,
namely $k=0,\ldots,k_{max}$, where $k_{max}=25$. Then we calculate
{\it analytically} $c_{\ell,m}$ from (\ref{relU}), $F$ from
(\ref{jrgf}), and $J^\theta$, $J^\varphi$ from (\ref{jthph}). For
the above $U$ we have
\be
F=-30 \rho^2 \sin \theta+\frac{35}{2} \rho^3 \sin \theta \cos
\theta \cos \varphi-\frac{35}{2} \rho^3 \sin^2 \theta \cos 2 \varphi.
\label{fex}
\ee
The analytical and the numerical values of $J^\theta$ and
$J^\varphi$ in the various $\theta_i$, $\varphi_j$ and $\rho_k$
are almost the same (the absolute value of their difference is of
order $10^{-7}$, at most).

We have also verified the validity of equation (\ref{gese3}) as
follows: We take $F$ from (\ref{fex}) and evaluate numerically
$U_a$ from (\ref{gese3}); $|U-U_a|$ is of order $10^{-5}$, at most.

\vskip 0.3cm
\noindent
{\bf Example 2}
\newline
\noindent
Let $U$ be given by
\be
U = \frac{1}{4 \pi} \frac{p_1 x_1+p_2 x_2+p_3 (x_3-a)}
{[{x_1}^2+{x_2}^2+(x_3-a)^2]^{3/2}}
\label{exampU}
\ee
with $a=0.5$ and $(p_1,p_2,p_3)=(0.1,-0.2,0.6)$. We evaluate $U$
for $\rho=1.5$ and the same equally spaced $\theta_i$ and
$\varphi_j$, as in Example 1. We again calculate numerically the
coefficients $c_{\ell,m}$ and then $U_a$.

For this example we found that the best value for $l_{max}$ is
10. For this value the difference $|U-U_a|$ is of order $10^{-6}$, at
most.

Finally, in Figure 1, we present the density plots of the
minimizing current $(J^\theta)^2+(J^\varphi)^2$ for the above
function $U$ in various cuts perpendicular to the $x_3$--axis.

\begin{figure}[ht]
\begin{center}
\epsfig{file=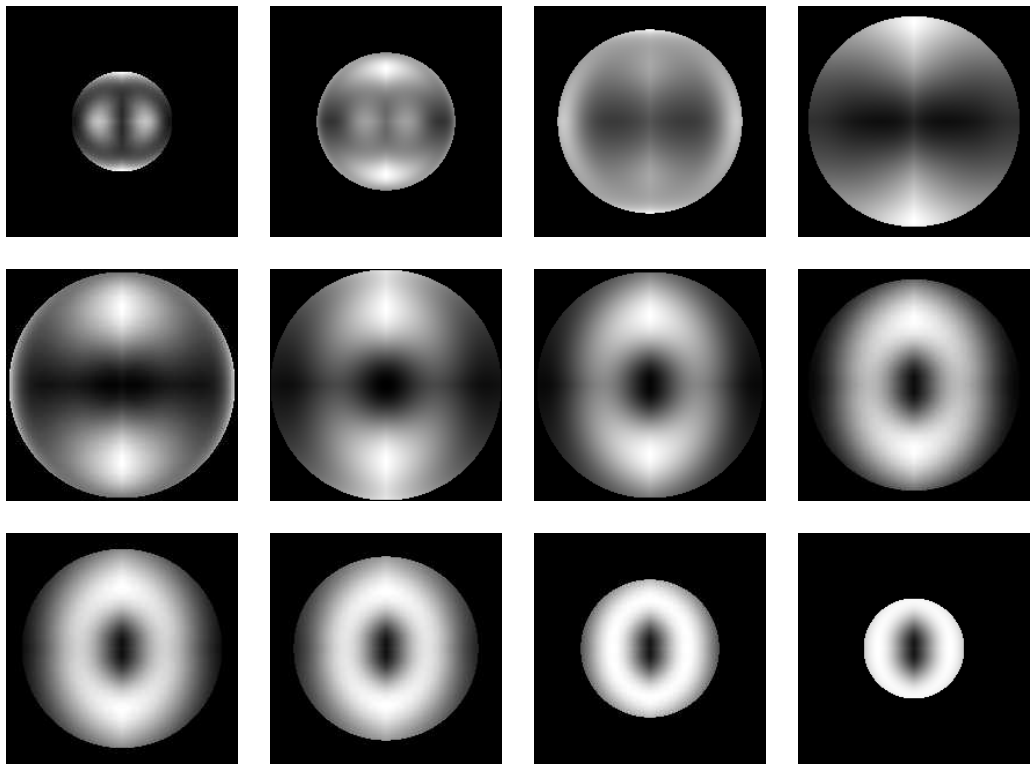}
\caption{Density plots for the minimizing current of the function
$U$ given by (\ref{exampU}). Starting from top left $x_3=-0.9$,
$-0.8$, $-0.6$, $-0.4$, $-0.2$, 0, $0.2$, $0.4$, $0.5$, $0.6$, $0.8$
and $0.9$.}
\end{center}
\end{figure}

\section*{Appendix A}
\setcounter{equation}{0}
\renewcommand{\theequation}{A.\arabic{equation}}

\noindent
{\bf Lemma.} Let
\be
U({\bf x},{\bf z}) \doteqdot \left( \frac{{\bf J}({\bf z}) \wedge
{\bf z}}{|{\bf x}-{\bf z}|(|{\bf x}| |{\bf x} - {\bf z}| +
{\bf x} \cdot ({\bf x}-{\bf z}))} \right) \cdot {\bf x},
\label{aU}
\ee
Then
\be
\int_{\mathbb{R}^3} (\Delta_{\bf x} U({\bf x}, {\bf z})) \Phi({\bf x})
d {\bf x} = -\frac{4\pi}{|{\bf z}|} \int^{|{\bf z}|}_0
\{ ({\bf J}({\bf z}) \wedge {\bf x}) \cdot (\nabla_{\bf x} \Phi({\bf x}))
\}_{{\bf x} = |{\bf x}| \frac{{\bf z}}{|{\bf z}|}} d|{\bf x}|,
\label{arel}
\ee
where $\Delta$ is the Laplacian (i.e.\ $\Delta = \nabla \cdot
\nabla$), and $\Phi({\bf x}) \in C^\infty_0 (\mathbb{R}^3)$.
\newline
\noindent
{\bf Remark}. As $\Delta U$ is singular close to
${\bf x} = |{\bf x}| \frac{{\bf z}}{|{\bf z}|}$, the integral in the
lhs of (\ref{arel}) should be understood in the sense of distributions.

\vskip 0.3cm
\noindent
{\bf Proof.}
Let ${\bf z}$ be at distance $a$ from the origin along the
direction $x_3'$.  Let $\Omega_\epsilon ({\bf z})$ denote a
small neighborhood of the interval $[0,{\bf z}]$ defined as follows,
\[ \Omega_\epsilon({\bf z}) = C_\epsilon ({\bf z}) \cup
S_\epsilon(0) \cup S_\epsilon ({\bf z}), \]
where $C_\epsilon({\bf z})$ is the cylindrical region
\[ C_\epsilon({\bf z}) = \left\{ {\bf x'} \in \mathbb{R}^3: \quad
\rho = \sqrt{{x_1'}^2+{x_2'}^2}=\epsilon, \quad 0 \le x_3' \le a \right\}, \]
while $S_\epsilon(0)$ and $S_\epsilon({\bf z})$ are the semi spherical
regions
\[ S_\epsilon(0) = \left\{ {\bf x'} \in \mathbb{R}^3 : \quad |{\bf x'}|
= \epsilon, \quad x_3'<0 \right\}, \]
and
\[ S_\epsilon({\bf z})=\left\{ {\bf x'} \in \mathbb{R}^3 : \quad
|{\bf x'} - {\bf z}|=\epsilon, \quad x_3'>a \right\}, \]
respectively.

Let $\Phi({\bf x})$ be a test function, then from the theory of
distributions it follows that
\begin{eqnarray}
\Delta U(\Phi) = \int_{\mathbb{R}^3} (\Delta U({\bf x}, {\bf z}))
\Phi({\bf x}) d {\bf x} & \doteqdot & \int_{\mathbb{R}^3} U({\bf x}, {\bf z})
\Delta \Phi({\bf x}) d {\bf x} = \nonumber \\
\lim_{\epsilon \to 0} \int_{\mathbb{R}^3/\Omega_\epsilon({\bf z})}
U({\bf x},{\bf z}) \Delta \Phi({\bf x}) d {\bf x} & = &
-\lim_{\epsilon \to 0} \int_{\partial \Omega_\epsilon
({\bf z})} \left( U \frac{\partial \Phi}{\partial n}-\frac{\partial
U}{\partial n}\Phi \right) dS, \label{alim}
\end{eqnarray}
where $dS$ denotes the infinitesimal surface element on the surface
$\partial \Omega_\epsilon({\bf z})$, $n$ denotes the unit outward
normal, and we have used the fact that $\Delta U=0$ in
$\mathbb{R}^3/\Omega_\epsilon({\bf z})$.
Let $I_1({\bf z}, \epsilon)$, $I_2({\bf z}, \epsilon)$,
$I_3({\bf z}, \epsilon)$ denote the contributions from the integration
along $C_\epsilon({\bf z}), S_\epsilon(0)$, and $S_\epsilon({\bf z})$,
respectively. It is easy to show that
$\mathop{\lim}\limits_{\epsilon\to 0}I_2 =
\mathop{\lim}\limits_{\epsilon\to 0}I_3 = 0$. We now compute $I_1$: Let
\[ f({\bf x'},{\bf z}) \doteqdot |{\bf x'}-{\bf z}|
(|{\bf x'}|| {\bf x'}-{\bf z}| + {\bf x'} \cdot ({\bf x'}-{\bf z})). \]
Thus if ${\bf x'} \in C_\epsilon({\bf z})$,
\[ f = \left[ (a-x_3')^2+\rho^2 \right] \sqrt{\rho^2+{x_3'}^2} +
\sqrt{\rho^2+(a-x_3')^2}(\rho^2+{x_3'}^2-ax_3'). \]
Hence
\begin{eqnarray*}
\frac{\partial f}{\partial \rho} & = &
2\rho\sqrt{\rho^2+{x_3'}^2}+\frac{\rho}{\sqrt{\rho^2+{x_3'}^2}}
\left[ \rho^2+(a-x_3')^2 \right] \\
& + & 2\rho\sqrt{\rho^2+(a-x_3')^2}+
\frac{\rho(\rho^2+{x_3'}^2-ax_3')}{\sqrt{\rho^2+(a-x_3')^2}},
\end{eqnarray*}
and
\[ \frac{\partial^2 f}{\partial
\rho^2}=2\sqrt{\rho^2+{x_3'}^2}+\frac{\rho^2+(a-x_3')^2}
{\sqrt{\rho_2+{x_3'}^2}}+2
\sqrt{\rho^2+(a-x_3')^2}+\frac{\rho^2+{x_3'}^2-ax_3'}
{\sqrt{\rho^2+(a-x_3')^2}}+\rho \tilde f, \]
where $\tilde f$ is bounded at $\rho = 0$. Evaluating  $f$, $\frac{\partial
f}{\partial \rho}$, and $\frac{\partial^2 f}{\partial \rho^2}$ at $\rho = 0$
we find
\[ f \bigg|_{\rho = 0} = x_3'(a - x'_3)^2 + (a - x_3') ({x_3'}^2
-ax'_3) = 0, \]
\[ \frac{\partial f}{\partial \rho} \bigg|_{\rho = 0} = 0, \]
\be
\frac{\partial^2 f}{\partial \rho^2} \bigg|_{\rho = 0} = 2x_3' + \frac{(a -
x'_3)^2}{x_3'} + 2(a-x_3') + \frac{{x_3'}^2 - ax_3'}{a -
x_3'}=\frac{a^2}{x_3'}.
\label{ader}
\ee
The integral (\ref{alim}) involves $-U\frac{\partial \Phi}{\partial \rho}+\Phi
\frac{\partial U}{\partial \rho}$. Also, since
\[ {\bf x'}=(x'_1, x'_2, x'_3), \quad {\bf z}=(0, 0, a),
\quad {\bf J} = (J_1, J_2, J_3), \]
it follows that
\[ ({\bf J}({\bf z})\wedge {\bf z}) \cdot {\bf x'} = a(J_2 x'_1-x'_2 J_1) =
a\rho(J_2\cos\varphi'-J_1\sin\varphi'), \]
where we have used $x'_1=\rho\cos\varphi'$ and $x'_2 = \rho\sin\varphi'$.

Equations (\ref{aU}) and (\ref{alim}) imply that we need to compute
\be
\lim_{\rho\to0} \int_0^{2 \pi} \int_0^{|{\bf z}|=a} \rho d \varphi'
d x_3' [a(J_2\cos\varphi'-J_1\sin\varphi')]
\left\{ -\frac{\rho}{f} \frac{\partial \Phi}{\partial \rho}+\Phi \left(
\frac{1}{f}-\frac{\rho}{f^2} \frac{\partial f}{\partial \rho} \right) \right\}.
\label{afinal}
\ee
However,
$\Phi({\bf x'}) = \Phi(\rho\cos\varphi', \rho\sin\varphi', x'_3),$ thus as
$\rho\to0$,
\[ \Phi = \Phi(0, 0, x'_3) + \rho\cos\varphi'\frac{\partial\Phi}{\partial x'_1}
(0, 0, x'_3) + \rho\sin\varphi'\frac{\partial\Phi}{\partial x'_2}(0, 0, x'_3)
+ 0(\rho^2), \]
and
\[ \frac{\partial\Phi}{\partial\rho}=\cos\varphi'\frac{\partial\Phi}{\partial
x'_1}(0, 0, x'_3) + \sin\varphi'\frac{\partial \Phi}{\partial x_2'}(0, 0,
x'_3) + 0(\rho). \]
Substituting the expressions for $\Phi$ and for
$\frac{\partial\Phi}{\partial\rho}$
in (\ref{afinal}), it follows that the rhs of equation (\ref{alim}) involves
\[ \lim_{\rho\to0} \int^{2\pi}_0 \frac{a\rho^3}{f^2}\frac{\partial f}{\partial
\rho} \left( J_1\sin^2\varphi'\frac{\partial\Phi}{\partial x'_2}
-J_2\cos^2\varphi'\frac{\partial\Phi}{\partial x'_1} \right)
d \varphi'= a\pi \left( J_1\frac{\partial\Phi}{\partial x_2'}
-J_2\frac{\partial\Phi}{\partial x'_1} \right) \lim_{\rho\to 0}
\frac{\rho^3}{f^2}\frac{\partial f }{\partial\rho}. \]
But
\[ \lim_{\rho\to 0} \frac{\rho}{f}\frac{\partial f}{\partial
\rho}=\lim_{\rho\to
0} \frac{\frac{\partial f}{\partial
\rho}+\rho\frac{\partial^2f}{\partial\rho^2}}{\frac{\partial f}{\partial
\rho}}=1+\lim_{\rho\to 0} \frac{\frac{\partial^2 f}{\partial
\rho^2}+\rho\frac{\partial^3 f}{\partial \rho^3}}{\frac{\partial^2
f}{\partial \rho^2}}=2. \]
Also
\[ \lim_{\rho\to 0} \frac{\rho^2}{f}=\lim_{\rho\to 0}
\frac{2}{f_{\rho\rho}} = \frac{2 x_3'}{a^2}. \]
Thus,
\[ \lim_{\rho\to 0} \frac{\rho^3}{f^2}\frac{\partial
f}{\partial\rho}=4\lim_{\rho\to 0} \frac{1}{f_{\rho\rho}}=4\frac{x'_3}{a^2}, \]
where we have used (\ref{ader}). Hence
\[ \lim_{\epsilon\to 0} I_1 = \frac{4\pi}{a}\int^a_0 \left[
J_1({\bf z}) \frac{\partial\Phi}{\partial x'_2} (0, 0, x'_3)-
J_2({\bf z}) \frac{\partial \Phi}{\partial x'_1}(0, 0, x_3')
\right] x'_3 dx_3'. \]

In the above derivation we have used the convenient set of
coordinates ${\bf x'}$, such that ${\bf z}$ is along $x_3'$. This
result can be immediately generalized by writing $I_1$ in an
invariant form. Then (\ref{arel}) follows.

\section*{Appendix B}
\setcounter{equation}{0}

We will show that $J^\theta$ and $J^\varphi$ can be expressed by
equation (\ref{jthph}). Indeed, if
\[ {\bf J}=J^\theta {\bf e}_\theta + J^\varphi {\bf e}_\varphi, \]
then the corresponding 1--form on the sphere of radius $\rho$ is
\[ \alpha^\theta d\theta +\alpha^\varphi d\varphi; \quad J^\theta =
\frac{1}{\rho} \alpha^\theta, \quad J^\varphi = \frac{1}{\rho \sin
\theta} \alpha^\varphi. \]
On a compact Riemannian manifold, any 1--form $\alpha$ has the unique
decomposition
\[ \alpha = dG + (-1)*d*\beta + \alpha^h, \]
where $G$ is a function, $\beta$ is a 2--form, $\alpha^h$ is a harmonic
1--form, and $*$ is the Hodge operator. Also there do not exist any nonzero
harmonic 1--forms on the sphere. Furthermore, $*\beta=F$, where $F$ is
a function. Hence
\[ \alpha = dG+(-1)*dF.\]
Using
\[ dG = \frac{\partial G}{\partial\theta} d\theta +
\frac{\partial G}{\partial \varphi} d\varphi, \]
and
\[ *dF = \frac{1}{\sin\theta} \frac{\partial F}{\partial\varphi} d\theta -
\sin \theta \frac{\partial F}{\partial \theta} d\varphi, \]
we find
\[ \alpha^\theta = \frac{\partial G}{\partial\theta} - \frac{1}{\sin\theta}
\frac{\partial F}{\partial \varphi}, \quad \alpha^\varphi = \frac{\partial
G}{\partial \varphi} + \sin \theta \frac{\partial F}{\partial \theta}, \]
and equations (\ref{jthph}) follow.

{\bf Remark}. In the case of $\mathbb{R}^3$, the analogous
decomposition is given by Helmholtz theorem: Let ${\bf A} = A^x
{\bf i} + A^y {\bf j} + A^z {\bf k}$, where ${\bf i}$, ${\bf j}$,
${\bf k}$ are the unit vectors along the $x$, $y$, $z$ axis, be a
vector field in $\mathbb{R}^3$. Then there exists a function $G$
and a vector field ${\bf B} = B^x {\bf i} + B^y {\bf j} + B^z {\bf
k}$ such that ${\bf A} = \nabla G + \nabla \wedge {\bf B}$. A
relationship between the general decomposition and the one in
$\mathbb{R}^3$ can be established using the following facts: (i) A
differential 1--form $\alpha = \alpha^x \, dx+\alpha^y \, dy
+\alpha^z \, dz$ can be canonically identified with the vector
field ${\bf A}$, where $A^x=\alpha^x$, $A^y = \alpha^y$, $A^z =
\alpha^z$. (ii) In $\mathbb{R}^3$ the Hodge operator transforms a
differential 1--form $\alpha$ into the differential 2--form $\beta
= \beta^{xy}\,dxdy + \beta^{yz}\,dydz + \beta^{xz}\,dxdz$, where
$\beta^{yz} = \alpha^x$, $\beta^{xz}=-\alpha^y$, $\beta^{xy} =
\alpha^z$. (iii) There do not exist any nonzero harmonic 1--forms
in $\mathbb{R}^3$.

\section*{Acknowledgments}

\noindent
This is part of a project jointly undertaken by the authors, A.A.\
Ioannides, and I.M.\ Gel'fand. A.S.F.\ is grateful to A.A.\
Ioannides for introducing him to MEG and for numerous important
discussions. This research was partially supported by the EPSRC.
V.M.\ was supported by a Marie Curie Individual Fellowship of the
European Community under contract number HPMF-CT-2002-01597. We
are grateful to the two referees for several important remarks.

\end{document}